\documentclass[twocolumn,showpacs,amsmath,amssymb,aps,prb,floatfix]{revtex4-1}
\usepackage{graphicx}
\usepackage{dcolumn}
\usepackage{bm}
\usepackage{subfigure}
\usepackage[varg]{txfonts}
\usepackage{color}   
\usepackage{hyperref}
\hypersetup{
    colorlinks=true, 
    linktoc=all,     
    linkcolor=blue,  
    citecolor=blue}
\def\bmatrix{\begin{pmatrix}}
\def\ematrix{\end{pmatrix}}

\begin{document}

\title{Quantum phase transition in a multi-connected superconducting Jaynes-Cummings lattice}
\author{Kangjun Seo}
\author{Lin Tian}
\email{ltian@ucmerced.edu}
\affiliation{School of Natural Sciences, University of California, Merced, California 95343, USA}
\date{\today}

\begin{abstract}
The connectivity and tunability of superconducting qubits and resonators provide us with an appealing platform to study the many-body physics of microwave excitations. Here we present a multi-connected Jaynes-Cummings lattice model which is symmetric with respect to the nonlocal qubit-resonator couplings. Our calculation shows that this model exhibits a Mott insulator-superfluid-Mott insulator phase transition at commensurate fillings, featured by symmetric quantum critical points. Phase diagrams in the grand canonical ensemble are also derived, which confirm the incompressibility of the Mott insulator phase. Different from a general-purposed quantum computer, it only requires two operations to demonstrate this phase transition: the preparation and the detection of commensurate many-body ground state. We discuss the realization of these operations in a superconducting circuit.
\end{abstract}
\maketitle

\section{Introduction\label{sec1}}
The past few years have witnessed stimulating progress in the study of superconducting quantum devices~\cite{squbitReview1,squbitReview2,squbitReview3}. Quantum logic operations with fidelity exceeding 99.9\% and quantum error correction codes were recently realized~\cite{Barends:2014,Chow:2014,Reed:2012}. By experimenting with various designs of the superconducting qubits and resonators, decoherence times on the scale of several tens of microseconds have been achieved in both 3-dimensional and planar circuits~\cite{Rigetti:2012, Reagor:2013, Barends:2013}. In several designs, such as the Xmon qubit, one qubit can be simultaneously connected to multiple resonators and control wires, which significantly improves the scalability and tunability of the superconducting systems~\cite{Barends:2013, Chen:2014, Niskanen:2007, Niemczyk:2010, Srinivasan:2011}. In the aspect of detection, quantum-limited amplifiers were developed to conduct phase-sensitive measurement of the amplitude of the microwave field and test quantum coherence effects at the single-photon level~\cite{CLang:2013, Riste:2013}.

The technological advancements in superconducting devices provide us with an appealing platform to explore many-body correlations. Analog and digital quantum simulators~\cite{Lloyd:96, Cirac+Zoller:12} of the superconducting systems have been proposed for numerous many-body effects, including phase transitions in the quantum spin systems~\cite{YDWangPRB2007, Garcia-RipollPRB2008, Tian:2010, GammelmarkNJP2011, MarquardtPRL2013, GellerPRA2013, Zhang:2014, SolanoPRL2014}, topological effects~\cite{NoriPRA2010, JQY2011, GreentreePRL2012, Schmidt:2013}, electron-phonon physics~\cite{Mei:2013,Stojanovic:2014}, and even high-energy physics~\cite{Kapit:2013, Marcos:2013, PeropadrePRB2013}. The implementation of these simulators can help us understand many-body phenomena that are hard to solve with traditional condensed matter techniques. Given the connectivity and tunability of the superconducting devices, we can also construct many-body Hamiltonians that do not exist in the real world, but carry novel many-body correlations. One such model is the so-called coupled cavity array (CCA) model, which is composed of an array of cavities each connected to neighboring cavities. Each cavity couples to a nonlinear medium, such as a qubit or a number of impurity atoms. In the pioneer works of Refs.~\cite{Hartmann:2006, Greentree:2006, Angelakis:2007, RossiniPRL2007, Na:2008, MakinPRA2008, Hartmann:2008, SchmidtPRL2009, Koch:2009, PippanPRA2009}, it was shown that the CCA exhibits the Mott insulator (MI)-to-superfluid (SF) phase transition for cavity polaritons, due to its resemblance to the Bose-Hubbard (BH) model~\cite{Fisher:1989, Batrouni:1990, KuhnerPRB2000}. The CCA has been thoroughly compared to the BH model in Refs.~\cite{Greentree:2006, Koch:2009}. Experimental efforts towards realizing the CCA with superconducting devices have also been conducted~\cite{Hoffman:2011,Houck:2012}.

In this work, stimulated by recent experimental progress, we present a multi-connected Jaynes-Cummings (JC) lattice model that demonstrates quantum phase transition for cavity polaritons. This model is constructed with arrays of qubits and resonators, where each qubit is connected to multiple resonators by exploiting the unique connectivity of planar superconducting qubits. In contrast to the CCA~\cite{Hartmann:2006, Greentree:2006, Angelakis:2007}, there is no direct coupling between the resonators. Instead, the qubit-resonator couplings in this multi-connected model serve both as onsite Hubbard interaction and as photon hopping. By varying a control parameter, this system can make a transition from the MI phase to the SF phase at commensurate fillings, similar to the CCA and the BH models. More interestingly, as the parameter is varied further, it makes another transition back to the MI phase from the SF phase. The MI-SF-MI phase transition is due to the symmetry with respect to the left and the right qubit-resonator couplings. These predictions are confirmed by our calculation of the single-particle density matrix and the energy gap of a small lattice using the exact diagonalization method~\cite{Cullum:1985}. This method has been previously used to study the BH model~\cite{Elesin:1994} and the CCA~\cite{Hartmann:2006, Angelakis:2007, MakinPRA2008}, where it gives qualitatively correct predictions of the phase transitions. We also obtain phase diagrams of the multi-connected JC model in the grand canonical ensemble at zero temperature, which indicate the incompressibility of the MI phase and the closing of the energy gap in the SF phase~\cite{Sachdev}. Note that due to the limitation of the current numerical method, details of the phase boundaries in the thermodynamic limit cannot be accurately characterized. One advantage of this system, compared with a general-purpose quantum computer~\cite{Lloyd:96}, is that it only requires two operations to demonstrate the phase transition: preparation and detection of the many-body ground state at commensurate fillings, both of which can be realized with current technology. 

Compared with previous works on the CCA~\cite{Hartmann:2006, Greentree:2006, Angelakis:2007}, our work exploits the nonlocal nature of the qubit-resonator couplings as well as the intrinsic symmetry with respect to the left and the right couplings to study quantum phase transition of cavity polaritons. This multi-connected JC lattice can be extended to two-dimensional or more complicated configurations to study many-body correlations in bosonic systems. The nonequilibrium dynamics of the cavity polaritons in this setup can also be investigated. We would like to mention that interconnected qubit-resonator arrays with uniform or opposite couplings were studied in previous works that focus on effective resonator coupling and quantum magnetism~\cite{QiuPRA2014, GarciaRipollPRL2014, GarciaRipollarXiv2014}. While our focus here is to study quantum phase transition caused by the interplay of the qubit-resonator couplings, which is distinctively different from that of the previous works. 

The paper is organized as follows. In Sec.~\ref{sec2}, we present the multi-connected JC lattice model and its construction with superconducting qubits and resonators. The effective Hubbard interaction and photon hopping are analyzed in the limiting case of drastically-different coupling constants. We calculate the single-particle density matrix and the energy gap of this multi-connected model at commensurate fillings using the exact diagonalization method in Sec.~\ref{sec3}. Then, in Sec.~\ref{sec4}, this method is extended to the grand canonical ensemble and the phase diagrams at zero temperature are derived. We discuss the realization of this model and the two operations required to demonstrate the MI-SF-MI phase transition: state preparation and detection. Conclusions are given in Sec.~\ref{sec6}.

\section{Multi-connected JC Lattice\label{sec2}}
\subsection{Model Hamiltonian\label{subsec2a}}
\begin{figure}
\includegraphics[width=0.9\linewidth,clip]{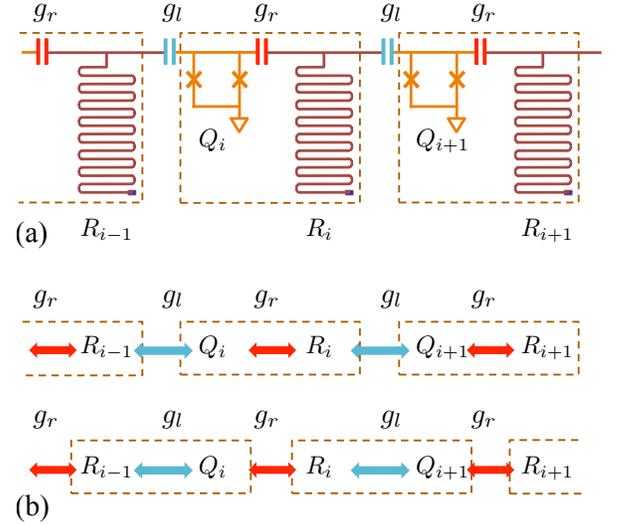}
\caption{(a) Schematic circuit of a multi-connected JC lattice with qubits $Q_i$, resonators $R_i$, and qubit-resonator couplings $g_{l}$ and $g_r$.  (b) Two ways of defining the unit cell: with $Q_{i}$ and $R_{i}$ in one cell (top) and with $Q_{i}$ and $R_{i-1}$ in one cell (bottom), respectively.}
\label{fig1}
\end{figure}

A 1D multi-connected superconducting JC lattice is depicted in Fig.~\ref{fig1} (a). This setup can also be extended to more complicated configurations, such as a two-dimensional checkerboard pattern of alternative qubits and resonators. The building block of this lattice is made of a superconducting qubit denoted by $Q_i$ and a superconducting resonator denoted by $R_i$. The qubit $Q_i$ couples to neighboring resonators $R_i$ and $R_{i-1}$ with coupling strengths $g_{r}$ and $g_{l}$, respectively. The total Hamiltonian of this model can be written as $H_{t} = \sum_{i} \left( H_{0}^{i} + H_{\text{int}}^{i} \right)$, where 
\begin{equation}
H_{0}^{i}=\omega_c a_i^\dagger a_i +  \frac{\omega_z}{2}\sigma^z_i\label{eq:H0i}
\end{equation}
is the noninteracting Hamiltonian of one repeating unit and 
\begin{equation}
H_{\text{int}}^{i}=g_r\left(a_i^\dagger \sigma_i^- + \sigma_i^{+} a_i \right)+g_{l} \left( a_{i-1}^\dagger \sigma_{i}^- + \sigma_{i}^+ a_{i-1} \right)\label{eq:Hinti}
\end{equation}
describes the JC couplings between a qubit and its neighboring resonators~\cite{CQEDtheory}. Here $\omega_c$ is the angular frequency of the resonator modes, $\omega_z$ is the energy level splitting of the qubits, $a_i$ ($a_i^\dagger$) is the annihilation (creation) operator of the resonator mode $R_i$, and $\sigma^{z,+,-}_i$ are the Pauli operators of the qubit $Q_{i}$. We set $\hbar = 1$ for convenience of discussion. 

The repeating units in our model are connected via qubit-resonator couplings. This is in sharp contrast to the CCA, where neighboring resonators couple directly to each other via a hopping Hamiltonian $-t \sum (a_i^\dagger a_{i+1} + a_i^\dagger a_{i+1})$~\cite{Hartmann:2006, Greentree:2006, Angelakis:2007}. As we will show, the qubit-resonator couplings in our model play both the role of onsite interaction and the role of photon hopping. A key feature of this model is that the system is invariant with respect to the exchange of the couplings $g_l$ and $g_r$. Hence, the unit cell can be defined in two ways, either with $Q_{i}$ and $R_{i}$ or with $Q_{i}$ and $R_{i-1}$ in one cell, as shown in Fig.~\ref{fig1} (b).

This multi-connected JC model can be realized with superconducting qubits and resonators developed in recent state-of-the-art experiments. One promising system is the so-called Xmon qubit, which excels in connectivity, controllability, and decoherence time~\cite{Barends:2013, Chen:2014}. This qubit can be connected to multiple resonators and control wires with tunable couplings. It also demonstrates a decoherence time exceeding $40\,\mu\textrm{s}$. In our discussions, we choose the control parameters to be in range of $g_{l,r}/2\pi \in[0,300]\,\textrm{MHz}$, the resonator detuning $\Delta/2\pi \in[-1,1]\,\textrm{GHz}$ with $\Delta=\omega_c - \omega_z$, and $\omega_{c}/2\pi=10\,\textrm{GHz}$. 

\subsection{Limiting case: $g_{l} \ll g_{r}$ (or $g_{r} \ll g_{l}$)\label{subsec2b}}
We start with the simple case of $g_{l}=0$, i.e., each repeating unit as defined by the top part of Fig.~\ref{fig1} (b) is isolated from each other with a vanishing coupling between $Q_i$ and $R_{i-1}$. Note that the opposite limit of $g_{r}\ll g_{l}$ can be studied similarly due to the symmetry between $g_l$ and $g_r$. The total Hamiltonian in this limit has the form of $H_{t} = \sum_{i} H_{JC}^{i}$ with 
\begin{equation}
H_{JC}^{i}=\omega_c a_i^\dagger a_i + \frac{\omega_z}{2} \sigma^z_i + g_r\left(a_i^\dagger \sigma^-_i + \sigma^+_i a_i \right).\label{eq:HJC}
\end{equation}
The Hilbert space of each unit cell is spanned by the basis states $\{|n_i,\sigma_i\rangle\}$ with $n_i$ being the microwave photon number of the resonator mode and $\sigma_i=\uparrow,\,\downarrow$ being the qubit state at site $i$. The lowest eigenstate of $H_{JC}^{i}$ is $|0_i,\downarrow_i\rangle $ with the energy $-\omega_z/2$. All other eigenstates, denoted by $|n_i,\pm_i\rangle$ with $n_{i}>0$, are polariton doublets in the subspace of $\{|n_{i}-1,\uparrow_i\rangle, |n_i,\downarrow_i\rangle\}$, and contain both photon and qubit excitations. The eigenenergies of the states $|n_i,\pm_i\rangle$ are $\varepsilon_{n_i,\pm_{i}} = ( n_i - 1/2)\omega_c \pm \Omega_{n_i} (\Delta)/2$ with $\Omega_{n_i} (\Delta)= \sqrt{\Delta^2 + 4 g_r ^2 n_i}$, depending on the detuning $\Delta$~\cite{CQEDtheory}. 

The qubit-resonator coupling $g_{r}$ generates nonlinearity in the polariton states. In Appendix A, we present an analysis of the nonlinearity involving only the lower-polariton states. The nonlinearity can be viewed as an effective Hubbard interaction for the polariton modes. Our results, different from that in Refs.~\cite{Greentree:2006, Koch:2009}, are in good agreement with the energy gap shown in Fig.~\ref{figA1}. For the low-lying states $|1_i,-_i\rangle$ and $|2_i,-_i\rangle$, the interaction strength $U=(2-\sqrt{2})g_{r}$ at $\Delta=0$; and $U=(\Delta+\vert\Delta\vert)/2$ for $\vert\Delta\vert\gg g_{r}$, demonstrating drastically-different behavior for large positive and negative detunings.

Next, we introduce a small but finite coupling strength $g_{l}$ that satisfies the condition $g_{l}\ll g_{r}$. This coupling can be viewed as a perturbation that induces hopping of a polariton excitation between adjacent unit cells with the conservation of the total excitation number, e.g., the nonzero matrix element
\begin{equation}
\langle 0_{i-1},\downarrow_{i-1}|\langle2_i,-_i|\sigma^+_i a_{i-1}|1_{i-1},-_{i-1}\rangle|1_i,-_i\rangle=-1/2\sqrt{2}\label{eq:hopping}
\end{equation}
is associated with the hopping of an excitation at site $i-1$ to site $i$ with a hopping strength $t\propto g_{l}$. 

The total Hamiltonian of the multi-connected JC lattice thus contains the two competing elements for a MI-to-SF phase transition~\cite{Sachdev}: onsite interaction and hopping between neighboring sites, both originated from the qubit-resonator couplings. With $g_{l}\ll g_{r}$ (or vice versa), the system is dominated by the onsite interaction and is expected to be in a MI phase at integer fillings. With the increase of $g_{l}$, the kinetic energy of the polariton mode eventually overcomes that of the Hubbard interaction, and the system could enter a SF phase. Given the symmetry between $g_{l}$ and $g_{r}$, these two couplings play similar roles when their strengths become comparable, each contributing to the onsite interaction as well as the hopping term. In the following sections, we will study the quantum phase transition of this model in detail. 

\section{Phase transition at commensurate fillings\label{sec3}}
Define the operator $\hat{N}=\sum_{i}(a_{i}^{\dag}a_{i}+\sigma_{i}^{+}\sigma_{i}^{-})$ as the total excitation number of the lattice, containing both photon and qubit excitations. Because $[H_{t}, \,\hat{N}]=0$, the total excitation number is a good quantum number. For a bosonic system, the MI phase occurs at commensurate fillings, i.e., the excitation number $N$ is a multiple of the lattice size $M$. Here we study the many-body phases of the multi-connected JC lattice with a fixed excitation number $N$ and $N/M$ being an integer. We apply the exact diagonalization method on a small lattice to find the precise ground state of this model~\cite{Cullum:1985}. This method gives qualitatively correct predictions of the phase transitions in the BH model~\cite{Elesin:1994} and the CCA~\cite{Hartmann:2006, Angelakis:2007, MakinPRA2008}. The natural choice of the basis vectors for our model is all possible configurations of the state $|\psi\rangle=|n_{1},\sigma_1\rangle |n_{2},\sigma_2\rangle \cdots |n_{M},\sigma_M \rangle$ that satisfies $\sum_{i}(n_{i}+\delta_{i})=N$, where $\delta_{i}$ refers to the qubit excitation at site $i$ with $\delta_{i}=0$ ($1$) for $\sigma_{i}=\downarrow$ ($\uparrow$). The Hamiltonian in the $N$-excitation subspace can be written as a sparse matrix on these basis vectors. Using a Lanczos-type algorithm, the low-lying eigenstates, in particular, the ground state, can be obtained. 

\subsection{Single-particle density matrix\label{subsec3a}}
For a system of fixed particle number, $\langle G\vert a_{i} \vert G \rangle\equiv 0$, and it cannot be utilized as an order parameter, where $\vert G\rangle$ is the many-body ground state. Instead, we calculate the normalized single-particle density matrix~\cite{Penrose:1956}
\begin{equation}
\rho_1(i,j) = \langle G | a_i^\dagger a_j | G \rangle/  \langle G | a_i^\dagger a_i | G \rangle,\label{eq:rho1}
\end{equation}
to characterize the phase transition of the multi-connected JC lattice. This matrix is generically Hermitian. Because of the lattice translational and reflectional invariances of the ground state, $\rho_1(i,j) = \rho_1(i+k,j+k)$ for an arbitrary integer $k$; and $\rho_1(i,j) = \rho_1(j,i)$. The matrix $\rho_1(i,j)$ is hence real, symmetric and cyclic. Below, we replace $\rho_1(i,j)$ by the notation $\rho_{1}(\vert i-j\vert)$. The single-particle density matrix, and hence the off-diagonal-long-range-order (ODLRO)~\cite{Yang:1962}, decays algebraically in the SF phase of 1D bosonic systems according to the Mermin-Wagner-Hohenberg theorem; whereas it decreases exponentially to zero in the MI phase. We can then choose a value of $x$, where $\rho_{1}(x)$ has significantly higher value in the SF phase than in the MI phase, and use $\rho_{1}(x)$ as a proof-of-principle indicator of the MI-to-SF phase transition, even though we cannot accurately determine the position of the quantum critical points. 

\begin{figure}
\includegraphics[width=0.95\linewidth,clip]{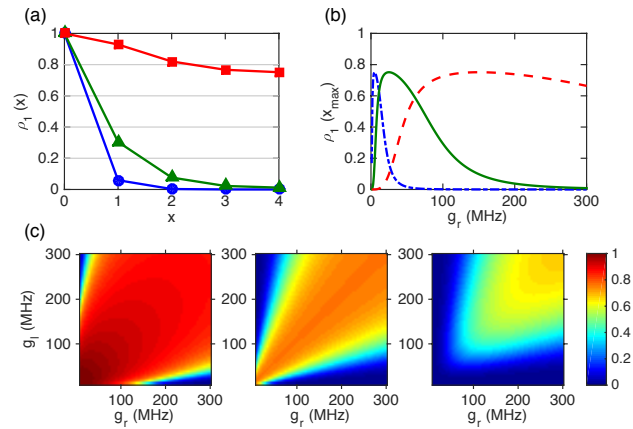}
\caption{(a) $\rho_{1}(x)$ versus the lattice distance $x$ at $\Delta = 0$. Blue circle: $(g_l,\,g_r)=(5,295)\,\textrm{MHz}$; green triangle: $(25,275)\,\textrm{MHz}$; and red square: $(150,150)\,\textrm{MHz}$. (b) $\rho_{1}(x_{\textrm{max}})$ versus $g_r$ at $\Delta = 0$. Blue dot-dashed curve: $g_l = 5\,\textrm{MHz}$; green solid: $g_{l}=25\,\textrm{MHz}$; and red dashed: $g_{l}=150\,\textrm{MHz}$. (c) $\rho_1(x_{\textrm{max}})$ versus $g_l$ and $g_r$ for $\Delta=-300,\,0,\,100\,\textrm{MHz}$ from left to right. Here $M=8$ and $N/M=1$.}
\label{fig2}
\end{figure}
We calculate $\rho_{1}(\vert i-j\vert)$ using the exact diagonalization method for a lattice of $M = 8$ and a total excitation number of $N=8$ under the periodic boundary condition. For such a lattice, the maximal lattice distance $x_{\textrm{max}}=4$. Our results show that even for a small-size system, this method can reveal the essential feature of the MI-to-SF phase transition. In Fig.~\ref{fig2} (a), $\rho_1(x)$ is plotted versus the lattice distance $x$ for three sets of couplings $(g_l,\,g_r)$. For $(g_l,\,g_r) = (5,295)\,\textrm{MHz}$, i.e., with $g_l \ll g_r$, $\rho_1(x)$ decreases to nearly zero as the lattice distance increases to $x=x_{\textrm{max}}$. This indicates that the system is in an insulator phase. As analyzed in Sec.~\ref{subsec2b}, in this limit, the coupling $g_{r}$ provides strong Hubbard interaction; while $g_{l}$ only induces small hopping. By slightly increasing $g_l$ to $25\,\textrm{MHz}$ and decreasing $g_r$ to $275\,\textrm{MHz}$, $\rho_1(x)$ increases but still nearly vanishes at $x=x_{\textrm{max}}$. In contrast, for $(g_l,\,g_r) = (150,150)\,\textrm{MHz}$, $\rho_1(x)$ remains finite at the maximal lattice distance $x_{\textrm{max}}$. Both couplings $g_{l,r}$ now generate hopping and onsite repulsion that are comparable in strength. The system hence demonstrates spatial correlation over a longer range than that in the MI phase, which implies the transition to a SF phase. 

The dependence of $\rho_1(x_{\textrm{max}})$ on the coupling $g_r$ is shown in Fig.~\ref{fig2} (b) for three values of $g_{l}$. For each $g_{l}$, $\rho_1(x_{\textrm{max}})$ decreases to zero when $g_{r} \ll g_{l}$ and $g_{r}\gg g_{l}$; and it reaches a large maximum when $g_{r}\sim g_{l}$. Hence, by continuously changing the coupling $g_{r}$ at a given $g_{l}$, the ground state evolves from a MI phase to a SF phase, and then makes another transition back to the MI phase. This is a unique feature of this multi-connected model, rooted in the symmetry with respect to the two couplings. In Fig.~\ref{fig2} (c), $\rho_1(x_{\textrm{max}})$ is plotted as functions of $g_{l,r}$ for three detunings, which further verifies the symmetry of the couplings. It also indicates that the detuning plays an important role in the phase transition. With a negative detuning, the system becomes more ``photon''-like with a reduced effective interaction, as discussed in Appendix A. The SF phase then becomes more favorable and exists in a broader parameter regime. With a positive detuning, on the other hand, the system becomes more ``spin''-like with a stronger effective interaction, and the SF regime is narrowed. 

\subsection{Energy gap\label{subsec3b}}
The energy gap is another important quantity to study the critical behavior of quantum phase transition. It is also related to the inverse of the compressibility of the many-body phases. Let $E_{N+} = E(N+1) - E(N)$ ($E_{N-} = E(N) - E(N-1)$) be the energy difference of adding (removing) one excitation to a system of $N$ excitations, where $E(N)$ is the ground state energy for a system with $N$ polaritons. The energy gap is defined as $E_{\textrm{gp}} = E_{N+} - E_{N-}$~\cite{Elesin:1994}. In the MI phase at commensurate fillings, $E_{\textrm{gp}}$ is finite due to the onsite interaction; while in the SF phase, $E_{\textrm{gp}}$ vanishes. 

\begin{figure}
\includegraphics[width=.95\linewidth,clip]{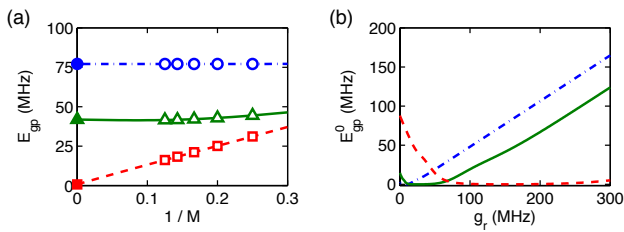}
\caption{(a) $E_{\textrm{gp}}$ versus $1/M$ at $M=4,5,6,7,8$ and its extrapolation $E^0_{\textrm{gp}}$ to $M=\infty$ at $g_r = 150\,\textrm{MHz}$. (b) $E^0_{\textrm{gp}}$ versus $g_r$. Here $g_{l} = 5\,\textrm{MHz}$ (blue circle, dot-dashed curve), $25\,\textrm{MHz}$ (green triangle, solid), and $150\,\textrm{MHz}$ (red square, dashed) with $\Delta=0$ and $N/M=1$.}
\label{fig3}
\end{figure}
We calculate the energy gap $E_{\textrm{gp}}$ of the multi-connected JC model at the filling factor $N/M=1$. In Fig.~\ref{fig3} (a), $E_{\textrm{gp}}$ is plotted as a function of $1/M$. Due to the finite-size effect, the energy gap remains open for a finite lattice in all regimes of the couplings. For $g_{r}\ll g_{l}$ or $g_{r}\gg g_{l}$, $E_{\textrm{gp}}$ is nearly independent of the size of the system; whereas for $g_r$ comparable to $g_l$, $E_{\textrm{gp}}$ strongly depends on $M$. We thus extrapolate the energy gap to the thermodynamic limit with $M\to\infty$ using a fourth-degree polynomial of $M$. The extrapolated gap $E^{0}_{\textrm{gp}}$, plotted in Fig.~\ref{fig3} (b) versus the coupling $g_{r}$ at fixed $g_{l}$'s, clearly bears the feature of a MI-to-SF phase transition. In the regime of $g_{r}\ll g_{l}$, where a MI phase is predicted, the gap $E^0_{\textrm{gp}}$ is open. With the increase of $g_{r}$, $E^0_{\textrm{gp}}$ decreases and eventually closes when $g_{r}$ becomes comparable to $g_{l}$, with this system entering a SF phase. As $g_{r}$ further increases towards $g_{r}\gg g_{l}$, $E^0_{\textrm{gp}}$ opens again after a finite interval of zero gap, indicating that the system is in the MI phase again. The energy gap in the limit of $g_{r}\ll g_{l}$ and $g_{r}\gg g_{l}$ can be well explained by a simple analysis of the effective onsite interaction, presented in detail in Appendix A and Fig.~\ref{figA1}.  

The above phase transition is featured by symmetric quantum critical points due to the symmetry between the couplings $g_{l}$ and $g_{r}$. At zero detuning, the many-body phase transition of this model is solely determined by the ratio $g_{r}/g_{l}$. For $g_{r}/g_{l}<\beta_{c}$ or $g_{r}/g_{l}>\beta_{c}^{-1}$ with $\beta_{c}$ being the critical point, the system is in the MI phase; and in the intermediate regime, the system is in a SF phase. From Fig.~\ref{fig3} (b), we estimate that $\beta_{c}\sim 2/3$. It can be shown that the phase transition at $\Delta\ne 0$ also embodies this feature. We want to mention that our numerical method, conducted on a small lattice, cannot yield accurate value for the critical points, which could change in the thermodynamic limit. Our results, however, demonstrate the main feature of the MI-SF-MI transition. 

\section{Phase Transition in Grand Canonical Ensemble\label{sec4}}
Quantum phase transition in the CCA is often studied in the grand canonical ensemble (GCE)~\cite{Hartmann:2006, Greentree:2006, Angelakis:2007}, where the excitation density (filling factor) is directly associated with the many-body phase and its compressibility. Here we extend the exact diagonalization method used in Sec.~\ref{sec3} to study the multi-connected JC lattice in the GCE~\cite{Cullum:1985}. Consider the free energy $\hat F = H-\mu \hat N$ at a given chemical potential $\mu$ and define $\vert G\rangle$ as the ground state of the free energy $\hat F$. In the GCE, the total excitation number $N$ is a function of the chemical potential, and can be obtained from the ground-state wave function by $N (\mu)=\langle G | \hat N |G\rangle$. The basis vectors in this calculation are: $\vert \psi\rangle = |n_{1},\sigma_1\rangle |n_{2},\sigma_2\rangle \cdots |n_{M},\sigma_M \rangle$ with $\sum_{i}(n_{i}+\delta_{i})\le N_{\textrm{max}}$ for a lattice of $M$ sites. The maximal total excitation number $N_{\textrm{max}}$ is chosen to include all possible basis vectors at the given chemical potential; and $N(\mu)\le N_{\textrm{max}}$. Note that the chemical potential, as discussed in previous works, is not a directly controllable parameter in this system~\cite{Koch:2009}.

\subsection{Excitation density\label{subsec4a}}
We calculate the many-body ground state of a lattice with $M=6$. The chemical potential is in a range that yields an excitation density of $n\in[0,\,2]$ with $n=N/M$. In Fig.~\ref{fig4} (a), the density $n$ is plotted as a function of the chemical potential at $\Delta = 0$. For the couplings $(g_l,\,g_r)=(5,295),\,(25, 275)\,\textrm{MHz}$, the density first increases with $\mu$ by small discrete steps of $\delta n=1/M$ to reach a broad plateau of $n=1$ at a critical chemical potential $\mu_{-}(n=1)$, as indicated by the solid circle. At $\mu \ge \mu_{+}(n=1)$, indicated by the solid square, the density starts increasing again to reach a plateau of $n=2$. The discreteness of the small steps is due to the finite size of this system, where the ground state always has fixed (integer) number of total excitations. The excitation number increases with the chemical potential one at a time, which gives the discrete density increment of $\delta n$. For $(g_l,\,g_r)=(150,150)\,\textrm{MHz}$, in contrast, no such plateau exists, and $n$ increases continuously with $\mu$ in small steps. These plateaus at commensurate fillings imply the incompressibility of the many-body state, which is an important feature of the MI phase~\cite{Sachdev}. The critical chemical potentials $\mu_{\pm}(n)$ correspond to the boundaries between commensurate and incommensurate densities, and hence, between the MI and the SF phases. The single-particle density matrix $\rho_1(x_{\textrm{max}})$ is plotted in Fig.~\ref{fig4} (b). When the chemical potential is within the plateaus, $\rho_1(x_{\textrm{max}})$ is reduced to a very small value (even in this finite size system), owning to the fast decay of the spatial correlation in the MI phase; whereas $\rho_1(x_{\textrm{max}})$ shows a slower decay outside the plateaus in the SF phase.  

\subsection{Phase diagrams\label{subsec4b}}
\begin{figure}
\includegraphics[width=1.0\linewidth,clip]{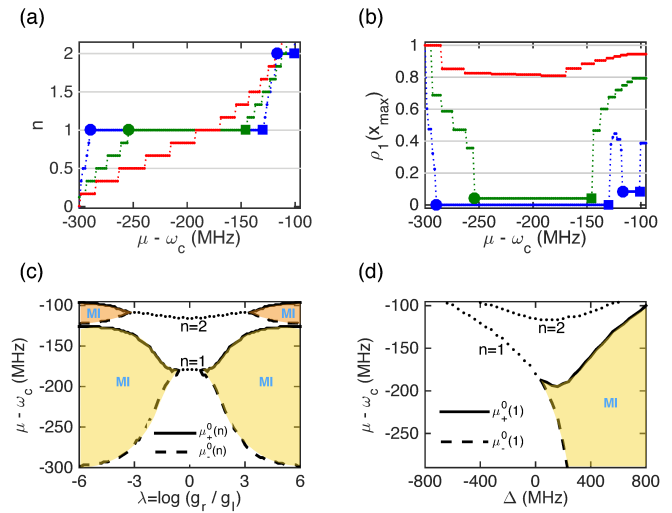}
\caption{(a) The density $n$ and (b) $\rho_1(x_{\textrm{max}})$ versus $\mu-\omega_c$ for a lattice of $M=6$ at $\Delta=0$. Blue dot-dashed curve: $(g_l,\,g_r)=(5,295)\,\textrm{MHz}$; green solid: $(25,275)\,\textrm{MHz}$; and red dashed: $(150,150)\,\textrm{MHz}$. The circles (squares) mark $\mu_{-}(n)$ ($\mu_{+}(n)$). (c) and (d) $\mu_{\pm}^{0}(n)$ versus $\lambda = \log (g_r/g_l)$ with $g_{r}+g_{l}=300\,\textrm{MHz}$ and $\Delta=0$ and versus $\Delta$ with $g_{l,r} = 150\,\textrm{MHz}$. Here $\mu_{+}^{0}(n)$ ($\mu_{-}^{0}(n)$) are solid (dashed) at the Mott lobes; dotted in the SF phase. Yellow (orange) lobes: $n=1$ ($n=2$).}
\label{fig4}
\end{figure}

The critical chemical potentials $\mu_{\pm}(n)$ discussed above define the phase boundaries for the transition between commensurate and incommensurate phases for the multi-connected JC lattice~\cite{Greentree:2006}. To derive the phase boundaries in the thermodynamic limit, we calculate $\mu_{\pm}(n)$ for finite lattices with $M=3,4,5,6$, respectively, and then extrapolate the results to $M\to\infty$ to derive $\mu_{\pm}^{0}(n)$. In  Fig.~\ref{fig4} (c), $\mu_{\pm}^{0}(n)$ are plotted versus the logarithmic ratio $\lambda = \log (g_r/g_l)$ with $g_r + g_l = 300\,\textrm{MHz}$ at $\Delta=0$ to form a phase diagram for our model. The regimes enclosed by $\mu_{\pm}^{0}(n)$ correspond to the Mott lobes at the commensurate fillings of $n=1,2$, demonstrating the incompressibility of the MI phase. As $\vert \lambda\vert$ decreases, $\mu_{+}^{0}(n) \to \mu_{-}^{0}(n)$, and the system exhibits a transition from the MI phase to the SF phase. Outside the Mott lobes, the dotted lines correspond to commensurate filling points within the SF regime~\cite{mupm}. The phase boundaries are symmetric with respect to positive and negative $\lambda$, due to the symmetry between the couplings. Furthermore, we plot $\mu_{\pm}^{0}(n)$ as a function of the detuning at $g_{l,r}=150\,\textrm{MHz}$ in Fig.~\ref{fig4} (d), which generates a phase diagram in the parameter space of $\mu$ and $\Delta$. Here the MI phase is more favorable at large positive detuning; while for $\Delta \lesssim 0.5\,\textrm{MHz}$, the system is always in the SF phase within the selected parameter range. These phase diagrams agree well with the results in Sec.~\ref{sec3} and our analysis in Appendix A. 

\section{Realization\label{sec5}}
In Sec.~\ref{subsec2a}, we briefly discussed the realization of the multi-connected JC lattice with superconducting qubits and resonators. Our model works in practical parameter regimes within reach of current technology. Recent experiments have shown that superconducting qubits can couple simultaneously to multiple resonators and control wires~\cite{Barends:2013, Chen:2014}. The detuning can be adjusted by applying dc field to tune the energy level splitting of the qubits. Tunable coupling in the qubit-resonator systems has been tested in several experimental works~\cite{Chen:2014, Niskanen:2007, Niemczyk:2010, Srinivasan:2011}. By varying one of the couplings (Fig.~\ref{fig3} (b)), the MI-SF-MI phase transition could be demonstrated. 

Compared with a general-purpose quantum computer~\cite{Lloyd:96}, this analog quantum simulator only requires two operations to be realized: 1. the preparation of the many-body ground state at selected control parameters and filling factor; 2. the detection of this ground state. Below we study the implementation of these operations and discuss the effects of quantum errors. 

\subsection{State preparation\label{subsec5a}}
The MI-SF-MI phase transition studied in Sec.~\ref{sec3} occurs in the ground state of the multi-connected JC lattice at integer fillings. We present a scheme to prepare the $N$-excitation ground state with $N/M=1$. This approach can be extended to prepare states with higher integer fillings. Our procedure contains  two steps: 1. flipping of the state of the superconducting qubits; 2. adiabatically transferring the system to the proper ground state using a Landau-Zener process. 

\begin{figure}
\includegraphics[width=1.0\linewidth,clip]{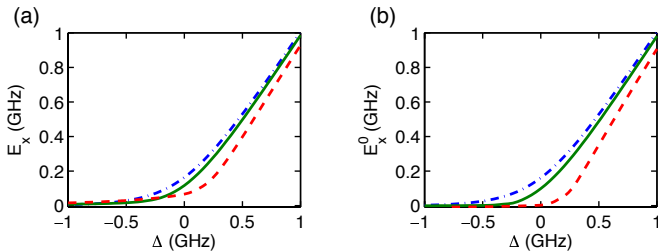}
\caption{(a) $E_{\textrm{x}}$ versus $\Delta$ for a lattice of $M=8$ and $N/M=1$ and (b) $E_{\textrm{x}}^{0}$ for $M\to\infty$. Blue circle: $(g_l,\,g_r)=(5,295)\,\textrm{MHz}$; green triangle: $(25,275)\,\textrm{MHz}$; and red square: $(150,150)\,\textrm{MHz}$.}
\label{fig5}
\end{figure}

We first discuss the excitation energy $E_{\textrm{x}}$ between the first excited state and the ground state of a lattice with $N=M$ excitations. The dependence of $E_{\textrm{x}}$ on the detuning is plotted in Fig.~\ref{fig5} (a) for a lattice of $M=8$. Here $E_{\textrm{x}}$ continuously increases with $\Delta$ and exhibits a linear dependence on $\Delta$ at large positive detuning. Due to the finite size effect, for $|\Delta|$ comparable to the couplings, the excitation energy remains sizable regardless of the many-body phase. When extrapolated to the thermodynamic limit with $M\to\infty$, however, $E_{\textrm{x}}$ is reduced to very small value in the regime of the SF phase and remains sizable for the MI phase, as shown in Fig.~\ref{fig5} (b). 

For state preparation, we first adjust the qubit energy to obtain a large positive detuning with $\Delta\gg g_{l},\,g_{r}$. Here the qubits are nearly decoupled from the resonators. The initial state of this system can be written as $|0_{1},\downarrow_{1}\rangle |0_{2},\downarrow_2\rangle \cdots |0_{M},\downarrow_M \rangle$ with $N=0$ excitation. By applying an ac driving field to generate a Rabi oscillation, the qubits are flipped to the state $|\uparrow_{i}\rangle$, and the system state becomes $|0_{1},\uparrow_{1}\rangle |0_{2},\uparrow_2\rangle \cdots |0_{M},\uparrow_M \rangle$. This state contains $N=M$ excitations and is the ground state of the multi-connected JC lattice in the limit of large positive detuning. Next, we adiabatically reduce the detuning to a target value, which is in a regime of interest to the study of the quantum phase transition. With the Landau-Zener theorem~\cite{LandauZener}, the final state is the many-body ground state at the target detuning. The time interval for the adiabatic process is determined by the excitation energy $E_{\textrm{x}}$, which remains a sizable value in all parameter regimes, e.g., $E_{\textrm{x}}=66\,\textrm{MHz}$ for $g_{l}=g_{r}=150\,\textrm{MHz}$ and $\Delta=0$, for a finite lattice of $M=8$. The state preparation can hence be implemented within tens of nanoseconds, much shorter than the decoherence time of the qubits and the resonators, and would not be seriously affected by the environmental noise. 

Because of the small anharmonicity in certain superconducting qubits, such as the transmon and the Xmon, the higher states in the qubit circuits can affect the state preparation scheme~\cite{Koch:07}. Let the third quantum state in a qubit be $|e_{i}\rangle$ and the energy level splitting between the states $|\uparrow_{i}\rangle$ and $|e_{i}\rangle$ be $\omega_{z}^{\prime}$. In a typical transmon (Xmon), the anharmonicity is $\sim5\%$ of $\omega_{c}$, yielding $(\omega_{z}-\omega_{z}^{\prime})/2\pi \sim 500\,\textrm{MHz}$. During the Rabi flipping, the ac field generates nonzero coupling between $|\uparrow_{i}\rangle$ and $|e_{i}\rangle$ which is of the same order of magnitude as the Rabi frequency $\Omega$ for the spin-flip operation. To avoid leakage to the state $|e_{i}\rangle$, it requires that $\Omega\ll (\omega_{z}-\omega_{z}^{\prime})$, which puts a constraint on the spin-flip time. By choosing $\Omega/2\pi=50\,\textrm{MHz}$, the spin flip can be realized in a practical time scale of $3\,\textrm{ns.}$.

\subsection{Detection\label{subsec5b}}
The phase transition can be characterized by measuring the quadrature correlation of the resonator modes at sites $i$ and $i+x_{\textrm{max}}$. Consider a quadrature component $X_{i}=a_{i}+a_{i}^{\dag}$ for the resonator mode $a_{i}$. The correlation of the quadratures $\langle X_{i}\cdot X_{j}\rangle $ can be detected by measuring the amplitude of the microwave field of both resonators and making a statistical average on the measured quadrature products. Such measurement has been utilized to study photon coherence and correlation in recent experiments~\cite{CLang:2013}. To achieve a faithful measurement of the many-body state, it requires that a single run during the measurement takes place in a time interval much shorter than the decoherence time of the qubits and the resonators. For a finite system with fixed number of excitations, $\langle a_{i}^{\dag}a_{j}^{\dag}\rangle \equiv 0$ and $\rho_{1}(i,j)$ is symmetric to $i$ and $j$. We then have $\langle X_{i} \cdot X_{i+x_{\textrm{max}}}\rangle = 2 \rho_1(x_{\textrm{max}})$. As discussed in Sec.~\ref{subsec3a}, $\rho_1(x_{\textrm{max}})$ carries the signature of the many-body phases and can be used to study the quantum phase transition. 

In addition, spatial correlation of the qubit operators also reveals the occurrence of the phase transition. We find that the correlation function $\langle \sigma_{i}^{+}\sigma_{j}^{-} + \sigma_{j}^{+}\sigma_{i}^{-}\rangle$ between the qubits at sites $i$ and $i+x_{\textrm{max}}$ demonstrates the same behavior as that of the single-particle density matrix presented in Sec.~\ref{sec3} and Sec.~\ref{sec4}. The phase transition can hence be detected by conducting measurements on the qubits. 

\section{Conclusion\label{sec6}} 
To conclude, stimulated by recent experimental progress in superconducting quantum devices, we studied the many-body phases of a multi-connected JC lattice model with nonlocal qubit-resonator couplings. We showed that a MI-SF-MI phase transition can be observed for cavity polaritons at commensurate fillings. Different from the CCA model studied in previous works, our model embodies a symmetry with respect to the qubit-resonator couplings, which is at the root of the appearance of symmetric quantum critical points. Our results for the single-particle density matrix and the energy gap confirm our analysis of an effective Hubbard interaction. Phase diagrams in the grand canonical ensemble are obtained, where the incompressibility of the MI phase is verified. We also studied the realization of this model with superconducting devices, presenting robust schemes for state preparation and detection. This model can be extended to two-dimensional qubit-resonator arrays and other more complicated configurations to study the many-body physics of microwave excitations. It also provides an interesting perspective to study the nonequilibrium dynamics of the cavity polaritons in this setup.  

\section*{acknowledgments}
This work is supported by the National Science Foundation under Award Number 0956064. L.T. thanks the Institute of Physics, Chinese Academy of Sciences, for hospitality. 
\vskip 3mm

\renewcommand{\thefigure}{A\arabic{figure}}
\setcounter{figure}{0}
\renewcommand{\theequation}{A\arabic{equation}}
\setcounter{equation}{0}
\renewcommand{\thetable}{A\arabic{table}}
\setcounter{table}{0}

\section*{Appendix A  Hubbard interaction in JC model\label{secappa}}
With $g_{l}=0$, the multi-connected JC lattice is an array of isolated qubit-resonator systems each described by the JC model. The qubit-resonator coupling generates nonlinearity in the JC model. We connect this nonlinearity to an effective Hubbard interaction for the polaritons with a simple analysis. 

The eigenstates $|n_i,\pm_i\rangle$ are the lower- and upper- polariton states with excitation number $n_{i}=\langle a_i^\dagger a_{i} +  \sigma_i^{+}\sigma_{i}^{-}\rangle$; and the state $|0_i,\downarrow_i\rangle $ contains no excitation. Note that the excitation number $n_{i}$ is a good quantum number in this model. We assume that the excitations fill the lower-polariton states only. Denote the energy to add $n_{i}$ excitations to this system as $\Delta\varepsilon_{n_{i}}=\varepsilon_{n_i,-_{i}} -\varepsilon_{0_{i},\downarrow_i}$. We derive 
\begin{equation}
\Delta\varepsilon_{n_{i}}=n_{i}\omega_{c}-\Delta/2-\Omega_{n_i} (\Delta)/2\label{eq:dEn}\\
\end{equation}
with $\Omega_{n_i} (\Delta)= \sqrt{\Delta^2 + 4 g_r ^2 n_i}$, using the expression for the eigenenergy in Sec.~\ref{subsec2b}. 

\begin{figure}
\includegraphics[width=1.0\linewidth,clip]{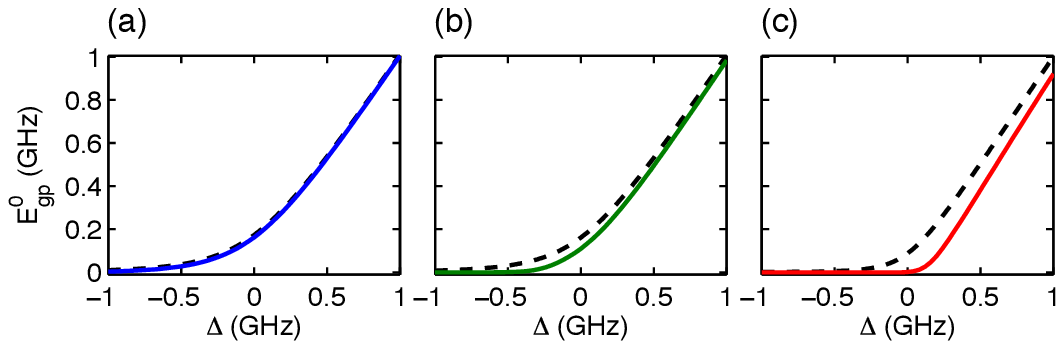}
\caption{$E_{\textrm{gp}}^{0}$ (solid) and effective Hubbard $U$ (dashed) versus $\Delta$. (a) $(g_l,\,g_r)=(5,295)\,\textrm{MHz}$; (b) $(25,275)\,\textrm{MHz}$; and (c) $(150,150)\,\textrm{MHz}$.}
\label{figA1}
\end{figure}
Assume that the lower-polariton states can be described by an effective Hamiltonian $H_{eff}=\omega_{p} p_{i}^{\dagger} p_{i}+ (U/2)p_{i}^{\dagger} p_{i}^{\dagger}  p_{i}p_{i}$, where $p_{i}$ is the annihilation operator of the polariton mode and $U$ is the strength of an onsite Hubbard interaction. Under this Hamiltonian, the energy of $n_{i}$ excitations is $\Delta\varepsilon_{n_{i}} = n_{i}\omega_{p} + Un_{i}(n_{i}-1)/2$. For $n_{i}=1$, $\Delta\varepsilon_{1_{i}}=\omega_{p}$. The effective interaction for $n_{i}$ and $n_{i}+1$ excitations can then be derived as $U=(\Delta\varepsilon_{n_{i}+1}-\Delta\varepsilon_{n_{i}}-\Delta\varepsilon_{1_{i}})/n_{i}$. Combining this result with Eq.~(\ref{eq:dEn}), we find the effective Hubbard interaction for the JC model as
\begin{equation}
U=\left[\Delta-\Omega_{n_i+1}(\Delta)+\Omega_{n_i}(\Delta)+\Omega_{1}(\Delta)\right]/2n_{i},\label{eq:U}
\end{equation}
depending on the coupling strength $g_{r}$, the detuning $\Delta$, and the excitation number $n_{i}$. For the low-lying states $|1_i,-_i\rangle$ and $|2_i,-_i\rangle$, which correspond to the lower-polariton states with one and two excitations, we have 
\begin{equation}
U=\frac{\Delta}{2}+\sqrt{\Delta^{2}+4g_{r}^{2}}-\frac{1}{2}\sqrt{\Delta^{2}+8g_{r}^{2}}.\label{eq:U12}
\end{equation}
This result is different from that in previous works using similar analysis~\cite{Greentree:2006, Koch:2009}.  

At $\Delta=0$, $U=(2-\sqrt{2})g_{r}$, determined by the coupling $g_{r}$. In the limiting case of $|\Delta|\gg g_{r}$, $U=(\Delta+\vert\Delta\vert)/2$, i.e.,
\begin{equation}
U=\{ \begin{array}{cl} 
0, & \Delta <0;\\
\Delta, &  \Delta > 0.
\end{array}
\end{equation}
For large negative detuning, the effective interaction vanishes. This is because the lower-polariton states in this regime are approximately photon-number states with equal energy level spacing. For large positive detuning, the interaction increases with the detuning. This offers us a convincing explanation of the behavior of the energy gap at the filling factor $N/M=1$. In Fig.~\ref{figA1}, we plot the effective interaction $U$ in comparison with the extrapolated energy gap $E^0_{\textrm{gp}}$ studied in Sec.~\ref{subsec3b}. In the regime of $g_{l}\ll g_{r}$ and $g_{r}\ll g_{l}$, the effective $U$ agrees very well with $E^0_{\textrm{gp}}$. This confirms the validity of our analysis for the effective interaction. 

We want to emphasize that this simple analysis only gives us a rough picture of the effective onsite interaction in the JC model, which decreases with the excitation number $n_{i}$. The JC model bears many properties that are distinctively different from that of the onsite Hubbard model.


\begin{thebibliography}{99}
\bibitem{mupm}The extrapolation method only gives approximate results for the critical chemical potential with $\mu_{+}^{0}(n)\approx\mu_{-}^{0}(n)$ in the SF phase. The difference between $\mu_{+}^{0}(n)$ and $\mu_{-}^{0}(n)$ is very small and cannot be distinguished from Fig.~\ref{fig4} (c,d).
\end{thebibliography}
\end{document}